\documentclass{camera}

\usepackage{graphicx}

\usepackage{natbib}
\usepackage{wrapfig}

\newcommand{\be}{\begin{equation}}
\newcommand{\ee}{\end{equation}}
\newcommand{\nn}{\mbox{} \nonumber \\ \mbox{} }
\newcommand{\ba}{\begin{eqnarray}}
\newcommand{\ea}{\end{eqnarray}}
\newcommand{\om}{\omega}

\newcommand\etal{\textit{et al.}}
\newcommand\eg{\textit{e.g.}}

\newcommand{\Bf}{{magnetic field}}
\newcommand{\Bfs}{{magnetic fields}}

\newcommand{\NS}{neutron star}

\begin{document}

\title{Gamma Ray Bursts:  back to the blackboard}

\author{Maxim Lyutikov}
\organization{Department of Physics, Purdue University, \\
 525 Northwestern Avenue,
West Lafayette, IN
47907-2036 }

\maketitle

\begin{abstract}
Exceptional observational breakthroughs in the field  of Gamma Ray Burst research are not paralleled by theoretical advances. In this review, based on the introductory  talk given at  the "The Shocking Universe" meeting, I argue that any  present day model of GRBs, especially of Short type, is grossly incomplete. I will highlight  various contradictions with observations that many models face and  briefly  mention a number of  ideas that might or might not work.   In particular, I will discuss (i) a possibility that early X-ray afterglows are coming from internal dissipation, and not from the forwards shock; (ii) that prompt radiation is beamed in the outflow frame.

\end{abstract}

\begin{quotation}
{\it 
It is impossible to obtain wages from a republic [...] without having duties attached. [...] so long as I am capable of lecturing and serving, no one in  the republic can exempt me from duty while I receive pay. I can hope to enjoy these benefits only from an absolute ruler.
}
\footnote{In modern words, Galilei did not want to teach and write grant applications in the Republic of Venice, thinking that only monarchy  in Florence can give him scientific freedom. As we know, years later he was proven wrong.}
\begin{flushright}
Galilei, {\it Opere x}, 348 ff
\end{flushright}

\end{quotation}

\section{Instead of introduction}

The launch of {\it Swift} satellite \citep{Swift} and ensuing discoveries, recent Fermi results \citep[\eg\ on GRB080916C][]{FermiGRB080916C}, combined with maturing of the fast response GRB observations at other wave lengths, are shaking the foundation of the GRB science \citep[for a recent review, see][]{GehrelReview}. The results of observations are  not conforming to simple expectations and are often contradictory to each other. At times like this, it is necessary to step back, to go back to the blackboard, 
and reconsider the basic issues in GRB physics. 
As we argue below, a long list of theoretical failures and contradictions with observations makes any present day model grossly incomplete.
The point of the  present review is to highlight these contradictions, and to re-evaluate again the fundamental principles of GRBs. It is not clear at the moment if these contradictions are fatal to the standard model or may be accommodated within  reasonable extensions (this, of course, also depends on a somewhat subjective definition of a ''standard model'').  I will briefly discuss a number of alternative ideas that might or might not work.

\section{Je pense donc je suis:
what are we sure about?}

Ren\'{e} Descartes argued that the only proof of our existence is that we think; all the rest is falsifiable. Asking a somewhat similar question, which parts of  GRB knowledge can we really be sure about? 

Observationally, we know that there exist {\it at least} two types of GRBs:  (i) Long-Soft bursts (referred to as {\it Longs} below), associated with young stellar populations, and typically associated with supernovae; (ii) Short-Hard bursts ({\it Shorts}), associated with old stellar population, not accompanied by SNe \citep{Kouveliotou93,Fruchter06}. The existence of two groups of bursts is a statistical statement: any given burst may be a Long in disguise of a Short, or vice versa, or otherwise "classificationally challenged" \citep{Perley} GRBs. For example, a number of Shorts show an extended emission tail, reminiscent of Longs (recall, also,  a  related possibility of a Li-Paczynski mini-SNÓ; \eg\ GRB 070724A \cite{Berger09}). A number of Longs do not show a SN signature \citep[\eg, GRB 060614][]{060614}, Longs occur outside of star formation \citep[\eg, GRB071003,][]{2006Natur.444.1050D,Perley}. 
In addition, the place of  X-ray Flashes (XRFs) and sub-luminous GRBs \citep[\eg][]{Soderberg06} in the classification of  extragalactic X-ray transients  is not clear: do  XRFs represent  a separate type of GRBs or just  precursors in regular GRBs?.

Theoretically, we are sure that GRBs come from the release of gravitational energy during  collapse of a stellar mass objects (a release of the rotational energy after the collapse also remains a distinct possibility). This certainty comes from the  association of  Longs with SNe. Shorts, in principle, can be anything, yet the fact that the host galaxies more often found than not for Shorts as well as for Longs,
makes us confident that they are also associated with astrophysical stellar-mass objects.  GRB explosions must be  strongly relativistic (from the compactness constraint) and most likely collimated (otherwise the total energy in some bursts would exceed $10^{55}$ ergs and it would be  hard to associate them  with an astrophysical 
object.

\section{Time scales}

\begin{figure}[h!]
 \begin{center}
\includegraphics[width=\linewidth]{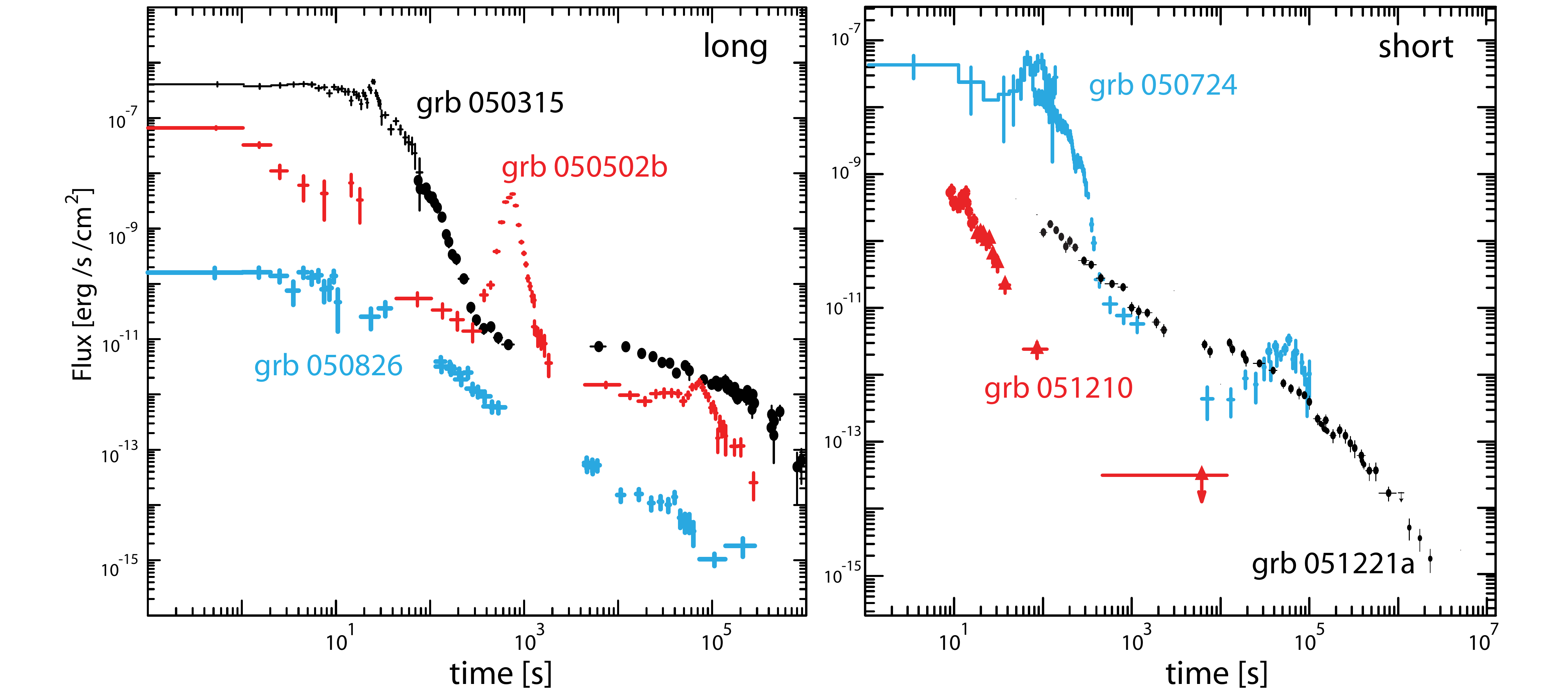}
\end{center}
\caption{Examples of X-ray afterglows of long and short  {\it Swift} events \citep{GehrelReview}. Note in particular flares in the short GRB050724 at $t\sim 100$ and $ 10^5$ sec. }
\label{GehrelsLightCurve}
\end{figure}

One of the difficulties in interpreting GRB light curves is the occurrence of various time scales, both  in the overall sample, as well as in many  separate burst, Fig. \ref{GehrelsLightCurve}. There is (i) the  time scale for the prompt emission, from sub-seconds to a few seconds for Shorts and up to thousand  seconds for Longs; (ii) X-ray  tails in Shorts lasting up to 100 seconds;  (iii) fast decay 
intervals in Longs, $t\sim 1000 $ sec, (iv) afterglow plateaus in Longs, $t\sim 10^3-10^4$ sec, (v) flares at the plateau phase and very late flares at times reaching $10^6$ sec; (vi) jet breaks at 
$10^3-10^5$ sec. Each of these times requires a separate explanation.  

Since GRBs are related to release of gravitational energy,  as a first approximation we can associated the time scale of the dominant energy release with an average density, 
$t \sim 1/\sqrt{G \rho}$ in the pre-GRB object. For Longs this gives $\rho \sim 10^3$ g cm$^{-3}$, for Shorts $\rho \sim 10^7$ g cm$^{-3}$. These are, surely just estimates, and they can be violated in both directions, meaning that under certain conditions the  energy release can occur on longer and even shorter time scales. 

In case of Longs, the density  $\rho \sim 10^3$ g cm$^{-3}$ corresponds to the central cores of massive stars. This relates naturally  to the SN-GRB connection \citep{smg+03}, which is, perhaps, the only  well established astronomical association for  GRBs. It is also natural to associate flares in Longs  at $ 10^3-10^4$ sec. (which can be significant in the total energy budget)  to  the average density  of the  He shell in pre-SNIb/c star. 

\begin{wrapfigure}{l}{0.6\textwidth}
 \begin{center}
  \vskip -.3  truein
\includegraphics[width=\linewidth]{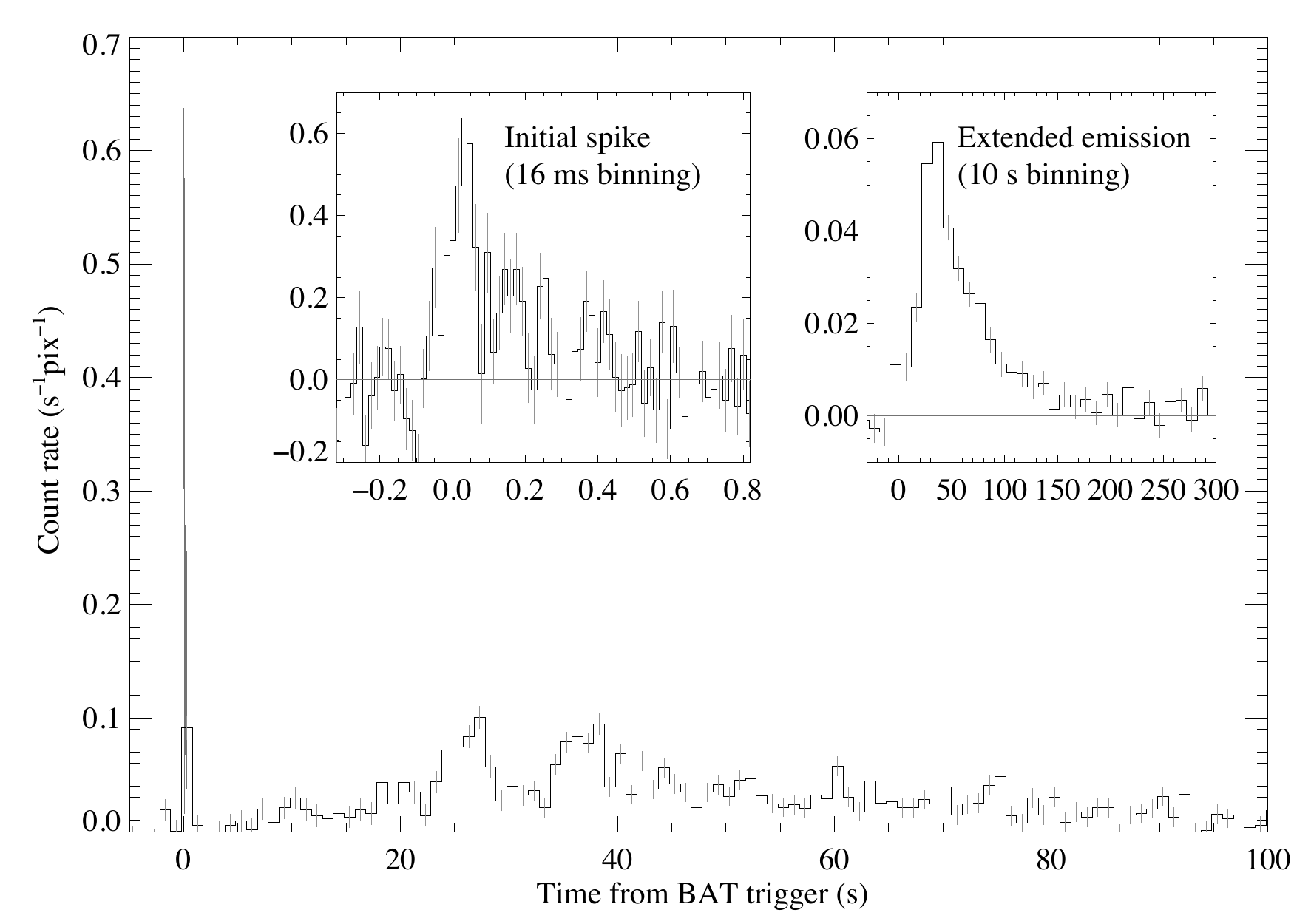}
\end{center}
 \vskip -.1  truein
\caption{The BAT light curve of a short GRB080503  \citep{Perley}. The fluence of the tail dominates over the prompt emission by a factor of 30. }
 \vskip -.1  truein
\label{ShortTail}
\end{wrapfigure}

Short bursts are more problematic, especially in view of the recent result that their fluence can be dominated by the extended tail \citep[][see also Fig. \ref{ShortTail}]{Perley}.  For prompt emission,  NS-NS  and NS-BH coalescence
  qualify. In addition,   in case of accretion induced collapse  (AIC)  of a white dwarf   \citep[\eg][]{YiBlackman},  the dynamical free fall time scale of the inside-out collapse might  be much longer than the energy release times scale at the moment of the bounce, when nearly \NS\ densities are reached, see \S \ref{23}.
 It is hardly reasonable to associate the time scales longer than $10^4$ sec with  the dominant release of gravitational  energy, but still it can be related to the  sub-dominant  energy release, \eg\ viscose time scale in an accretion disk \citep{CannizzoGehrels}.

If GRBs are powered by the release of rotational energy \citep[\eg][]{Usov92}, the rate of energy release is controlled by \Bf. The required fields are 
\be
B \sim  { c^{5/2} \over G \sqrt{M t}} \approx 5 \times 10^{16} \, {\rm G}\, \left( {t \over 1 {\rm sec}} \right)^{-1/2}
\ee
This is, perhaps, too high of a \Bf\  for Shorts, but can be important for Long prompt emission and early afterglows. 

\section{The standard model} 

The phrase "Standard model" means different things to different
people, so let us first define the terminology. GRBs release large amounts of energy, creating an outflows relativistically expanding into circumburst medium. In describing the bulk flow properties, the continuous fluid description is applicable  \citep[as opposed to kinetic model, advocated, \eg,  by][]{2003PhRvL..90h5001L}.
If the ejecta plasma is not too strongly  magnetized,  two shocks  form, the forward shock   (FS) in the circumburst medium and the reverse shock in the ejecta, separated by the contact discontinuity. In case of  strongly  magnetized ejecta the reverse shock may not be able to form; instead, a rarefaction wave would propagate from the contact into the ejecta. 

A particular variation of this general picture, which we will call ''the standard model"  \citep{MeszarosRees92,Sari95,PiranReview,MeszarosReview}  advocates that the prompt emission is generated in the baryon-dominated  ejecta through internal shocks, while the forward shock  (and  partially the reverse shock) produce long lasting broadband emission -- afterglow --  from  radio to X-ray bands. The spectral evolution of the synchrotron FS emissivity is well defined, with \Bf\ and particle energies simply parametrized with  the FS Lorentz factor.  
 The main impetus for this interpretation came from the fact that in  the BATSE/Beppo-SAX era, the afterglows preferentially showed smooth light curves, with power-law decays. This was
attributed to the emission of a forward shock. In a self-similar regime, the parameters of the FS evolve smoothly with time; FS is  also fairly insensitive to possible variation in the external density and/or late  sub-dominant luminosity  fluctuations of the central engine. In addition, the fact the afterglows of Shorts  and Longs look similar, is naturally explained: at  the self-similar stage, the details of the central engine are not important, only the total injected energy and outside density matter.

Though the overall hydrodynamic structure of the outflow consisting of the  FS, contact discontinuity and (possibly) a reverse shock is hardly questionable, the observational signatures of shock components are not clear at the moment. In the  {\it  Swift} era, the early afterglow light curves, at times $\leq 10^5$ sec, show numerous variability phenomena, like fast decays, plateaus, flares, various light curve breaks. This variability is  hard to explain within the FS model, which is expected to produce a smooth light curve. Also, the conventional RS signature, a  bright flash followed by  a smooth power law decay in the optical band, is rarely observed. 

Thus, as we discuss below, it is not clear if and when, and at which wave band the late time emission, which we call an afterglow, comes from the FS and whether or not it may come from internal dissipation. (We stress that we use the term afterglow in a relation to late, post-prompt, emission, not necessarily coming from the FS.

\section{Prompt emission mechanism}

One of the most exemplary illustration of   our lack of understating of GRBs, is that
after  years of intense  research we are still not sure about the most basic property of GRB, the  prompt emission mechanism. The demands on the mechanism are fairly basic: it should produce 
a non-thermal signal peaking at about $1 $ MeV (there are good reasons to believe that even after  the recent detections of bright GeV emission by {\it Fermi}, the very high energy emission is  a separate, and likely energetically subdominant emission mechanism.) The two main discussed possibilities for the prompt emissions are synchrotron and Inverse Compton (IC) processes.

\subsection{Synchrotron}

Perhaps the best argument for  synchrotron emission come from the fact that with some very  generic scaling  it reproduces the \cite{Amati} relation
(most others GRB correlations, especially those involving jet break times, are being strongly questioned in the  {\it Swift} era). 
If we assume that a fixed fraction $\epsilon _B$ of the central luminosity comes out in a form of \Bf, we find
\be 
\epsilon _B L \approx \Gamma^2 B^2 r^2 c,\, B \approx {1 \over \Gamma r}\sqrt{\epsilon _B L\over c} 
\ee
where $B$ is \Bf\ in the jet frame, $\Gamma$ is bulk Lorentz factor and other notations are standard.
If particles are accelerated (at the  shocks or in the reconnection layers) to random Lorentz factor $\gamma \sim m_p/m_e$ 
\citep[this is indeed seen in simulations of e-i shocks][but {\it not} in magnetized transverse shocks thought to  occur in GRBs]{2008ApJ...682L...5S}), the synchrotron  photon energy
\be 
E_p \approx \hbar \Gamma \gamma^2 {e B \over m_e c} = { e \hbar \over c^{3/2} m_e} {\gamma^2 \over r} \sqrt{\epsilon _B L}
\approx 100 {\rm keV} r_{14}^{-1} L_{52}^{1/2}
\label{Ep}
\ee
 falls approximately in the correct  observed range.
This immediately reproduces the \cite{Amati} relation {\it  independent of the bulk Lorentz factor} $\Gamma$  \citep[if emission radius is limited from below, $r>r_{min}$, Eq. (\ref{Ep}) gives an inequality][]{
2005MNRAS.360L..73N}.
In addition, the non-thermal  \citep{Band93} spectrum is natural in the  synchrotron model. 

On the other hand, the synchrotron model has a number of serious problems: (i) low frequency tails are wrong \citep{Preece98,Ghisellini00}; (ii) the dependence of the peak frequency  on the radius of emission is strong, and if one tries to relate the radius to the observed variability (in the framework of the standard model, $r \sim t_v \Gamma^2$), one regains dependence on $\Gamma$ and the predicted dependence between peak energy $E_p$ and variability is not confirmed. Also, if the prompt-GeV correlation reported by {\it Fermi} \citep{FermiGRB080916C} is taken to imply  large radii of prompt  emission, $r\sim 10^{16}$ cm, then  the  high peak energy for GRB080916C   is  inconsistent with Eq.  (\ref{Ep}).

The origin of \Bfs\ is also a question. They  can be amplified at  the shock (it is required that $\sim  10\% $ of energy goes into B-field; simulations typically do not show such a high level of turbulence, yet observations of young  SNRs, where FSs  are non-relativistic,  are consistent with such high fields  \cite{Reynolds08}). If \Bfs\ come from the central source, the observational  evidence, primarily polarization, are inconclusive \citep{CoburnBoggs,RutledgeGRB}, though  a recent detection of high polarization in optical (Steele \etal) is exciting.
Finally, \Bfs\ may be  amplified by dynamo in the bulk \citep[\eg\ due to Richtmyer-Meshkov and/or  RT instabilities][]{GoodmanRM}.

\subsection{Inverse Compton emission}

There are two types of models invoking IC effect: quasi-thermal comptonization and comptonization by relativistic leptons (e.g., SSC-type models).

\subsubsection{Quasi-thermal comptonization}

In most models of  GRBs,  thermal photons with rest frame temperature $  T \sim 2 \times 10^8 \, {\rm K } \approx 20$  keV   may carry a large fraction of energy at 
the photospheric radius \cite[this  temperature estimate comes from balancing the  rates of expansion 
and  pair-production][though a post-transparency acceleration might also be important]{Goodman86,Paczynski86}; observational  signatures of these thermal photons are discussed by  \cite{Ryde05}.  Boosting these photons to observer frame  will gives a peak at  $\sim$MeV range, approximately as observed. 
All one needs to do, is to scatter these photons off a weakly relativistic leptons to produce a non-thermal tail \citep{Thompson94,Ghisellini99,
Peer08,Beloborodov09} (thermal comptonization of the black body photons is not important  \cite{Beloborodov09}). 

These models have a number of  advantages, (i) strong thermal component is predicted in most models \citep[\eg][]{LyutikovUsov}; if it is not seen, then where it is? (ii) the model naturally explains hard low energy tails: it's a Rayleigh-Jeans tail, possibly modified by comptonization and by addition of emission from regions with different break frequencies. 

\subsubsection{Comptonization by relativistic leptons}

If the prompt emission is produced via IC scattering by a  population of relativistic electrons \citep[\eg][]{KumarPanaitescu08}, then  lots of low frequency seed photons  are needed; observations typically do not find these seed photons \citep[][though recent observations of GRB 080319B do show large optical fluxes, correlated with $\gamma$-rays]{PiranICGRB}.
In addition, these types of models often invoke Compton $Y$ parameter larger than unity, so that most energy comes out in higher Compton peaks, putting even more demands on the total energy. 
Comptonization by relativistic leptons is, perhaps, the best model for the very high energy emission detected by {\it Fermi}. 

\subsection{ Dissipation and acceleration mechanism}.

Regardless of the emission mechanism, we still do not understand the particle acceleration mechanism. Though shocks are nearly universally hypothesized to do the job, the
particle acceleration at shocks, especially relativistic, is far from 
understood and simple prescriptions should not be taken for granted. For example, modern PIC 
simulations of   magnetized transverse shocks do not show any acceleration \citep{2008ApJ...682L...5S} at the shock front.
(An alternative possibility  is that  particles may be accelerated downstream in an electron-ion plasma at the ion cyclotron resonance \cite{1992ApJ...390..454H,2006ApJ...653..325A}.)  Alternative possibilities are: (i) acceleration at magnetic reconnection \citep[\eg][]{Thompson94,LyutikovJPh}, which is even less understood than shock acceleration; (ii) collisional dissipation \citep{Beloborodov09} due to relative streaming of protons and neutrons in the flow (operating fairly close to the photosphere and producing quasi-thermal Comptonization with the Band spectrum). 

\section{Problems and contradictions}
\label{Problems}

Let us next discuss the most outstanding problems and contradictions in the field of  GRB research.  Many of the problems listed below are generic to the most of  present day models, while some are specific to the standard model. 

{\bf Early afterglow energetics. } A number of  early, $\leq 10^4 sec$ X-ray afterglows of Long GRBs show a  plateau phase, when the observed flux decays slowly, $F_\nu  \propto t^{-1/2}$. If associated with FS, this requires that the FS energy  $E_{FS}$ increases with time, $E_{FS} \propto t^{1/2}$. Thus, a central engine should inject $\sim 10$ times more energy at the plateau phase  than was available  during the prompt phase. Since the total energy in the outflow is constrained by late radio observations, this put unreasonable demands on efficiency of prompt emission.

\begin{wrapfigure}{l}{0.5\textwidth}
 \begin{center}
\vskip -.25 truein
\includegraphics[width=\linewidth]{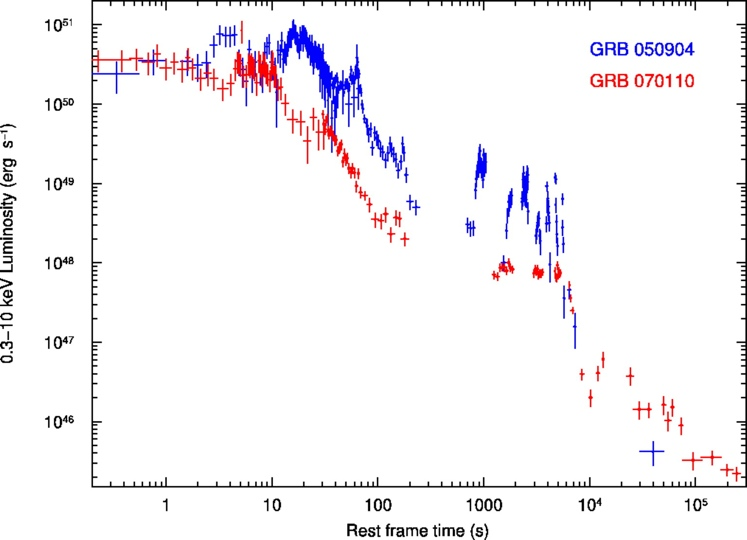}
\end{center}
\vskip -.3 truein
\caption{Combined BAT and XRT light curves of GRB050904 
and GRB 070110 in the source rest frame \citep{Troja07}. Drops after plateau  are sharper than from an instantaneous switch-off of relativistic outflow, emitting at $\geq 10^{13} $ cm.
 }
\label{Troja}
\end{wrapfigure}

{\bf Abrupt endings of the plateau phases}.
In a number of cases, the 
  plateaus end with an abrupt drop in luminosity 
 \citep[][Fig. \ref{Troja}]{Troja07}. Drops after plateau  are sharper than from an instantaneous switch-off of relativistic outflow, emitting at $\geq 10^{13} $ cm, and thus cannot be coming from a FS.
On the other hand, the spectrum is continuous over the break \citep{Nousek06}, indicating  that pre- and post-break emission mechanisms are similar. 
\,\,\,\, {\bf Production of flares: energy constraint.}
In the framework of FS afterglows, flares correspond to additional injection of energy into the FS. Each flare with the flux increase of the order of unity requires energy injection of the order of total energy already in the outflow. Thus, in case of GRBs with many flares,  the total energy in the FS grows in geometrical progression. 
In addition, post-flare evolution should correspond to an outflow with higher energy; this contradicts the fact that after the flare the flux typically returns to the extrapolated pre-flare level.

\begin{wrapfigure}{l}{0.5\textwidth}
 \begin{center}
  \vskip-.25 truein
\includegraphics[width=\linewidth]{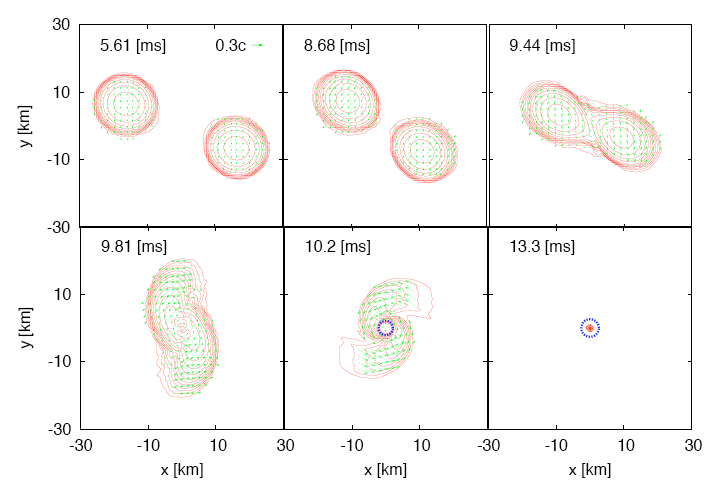}
\end{center}
\caption{Snap shots of  the density contours for NS-NS coalescence \citep{Kiuchi}. Note that the active stage of coalencence takes approximately 10 msec. 
 }
\label{KiuchiNSNS}
\end{wrapfigure}

{\bf Production of X-ray tails and   flares in Shorts: time constraint}
Particular problematic (from the theoretical point of view) are flares and tails in Shorts. 
 Numerical simulations indicate that the active stage of NS-NS coalescence typically takes 10-100 msec. A small amount, $\leq 0.1 M_\odot$  of material may be ejected during the merger and accretes on time-scales of 1-10 secs, depending on the assumed $\alpha$ parameter of the disk \cite[\eg][ Fig. \ref{KiuchiNSNS}]{Kiuchi, LiuNSNS,Faber}.   Thus, any energetically dominant  activity on much longer time scales  contradicts the NS-NS coalescence paradigm  for Short. 
\,\,\,\, Most puzzling are 
 cases when the tail fluence  dominates over the primary burst \citep[by a factor of 30 as in GRB080503][]{Perley}, or when powerful flares appear late in the afterglow (\eg,  in case of GRB050724 there is a powerful flare at $10^5$ sec, Fig. \ref{GehrelsLightCurve}). In the standard FS model of afterglows this requires that at the end of the activity, lasting 10-100 msec,  the source releases  more energy than during the prompt emission  in a form of a low $\Gamma$ shell, which collides with the FS  after $\sim 10^6$ dynamical times, a highly fine-tuned  scenario.

 {\bf Fast optical variations}
There are a number of GRBs that show fast optical variability (\eg\ GRB021004 and most notoriously GRB080916C). This is hard to explain within the standard FS-RS model, since optical electrons should be in a  slow cooling regime \citep[][suggested a model in which synchrotron emitting optical electrons are cooled by IC emission]{2009MNRAS.395..472K}.

{\bf Naked  GRBs}
Naked  GRBs ( GRBs without an appreciable afterglow emission, \eg, GRB050421) are hard to produce in the FS model \citep{KumarPanaitescuNakedGRB}. 

{\bf Closure relation}. 
Emission for a FS predicts a definite relation  between the 
 spectral and temporal indexes, $F_\nu \propto \nu^{-\beta} t^{-\alpha}$ \citep{ReesMeszaros98}. Observed afterglow typically do not comply with this prediction \citep[\eg][]{Racusin09}

\begin{wrapfigure}{l}{0.54\textwidth}
 \vskip - .3 truein
\includegraphics[width=\linewidth]{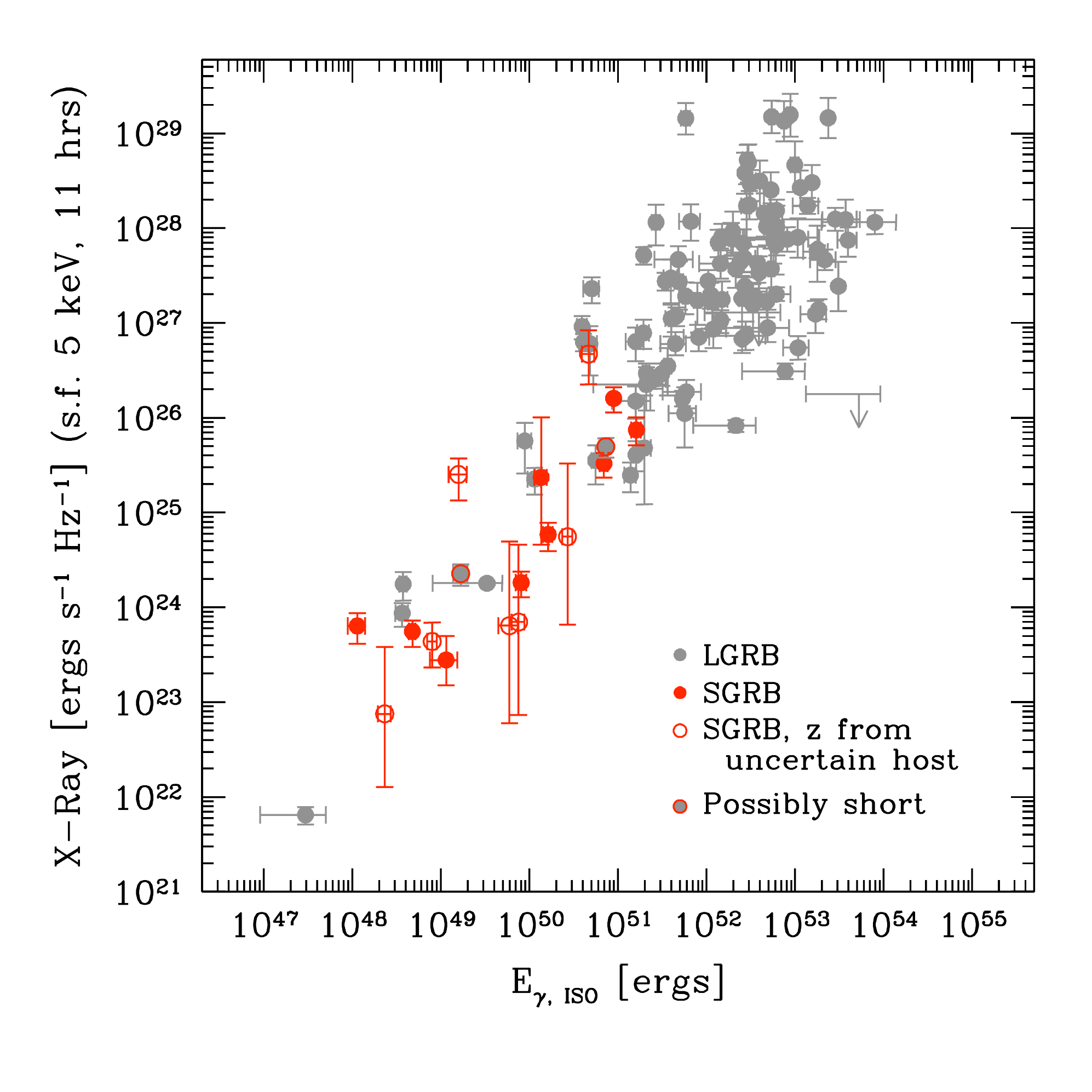}
\vskip - .3 truein
\caption{Isotropic-equivalent luminosity of GRB X-ray afterglows scaled to $ t  =11$ h ($5-keV$ source frame) after the burst trigger as a function of their isotropic $\gamma$-ray energy release \citep{Nysewander}. Note that Shorts and Longs form a continuous sequence. }
\label{LxLiso}
\end{wrapfigure}

 {\bf Similarity of Short's and Long's afterglows}.
Another puzzling fact is that early afterglows of Longs and Short look surprisingly similar, forming a continuous sequence, \eg, in relative intensity of X-ray afterglows as a function of prompt energy \cite[][Fig. \ref{LxLiso}]{Nysewander}.
 This is surprising in a forward shock model:  the properties of the forward shock do depend on the external density, while the prompt emission is independent of it.  The difference between circumburst media densities in Longs (happening in star forming regions) and Short
(happening in low density galactic or even extragalactic medium) is many orders of magnitude.  
In defense of the forward shock model, one might argue that afterglow dynamics depends on $E_{ejecta} /n$, both of which are orders of magnitude smaller for Shorts if compared with Longs. 
Yet afterglows are very similar and, most importantly, form a continuous sequence; this is surprising if 
afterglow luminosities are determined by the drastically different  circumburst density. 
 \\
\,\,\,\, {\bf (A)-chromatic breaks} In the standard FS model,   a so called jet break is  expected when the FS Lorentz factor becomes of the order of the jet opening angle \citep{Rhoads99}. Such light curve breaks should be achromatic. This contradicts  observation,  often indicating that  the breaks are chromatic, with different jumps in optical and X-rays, and sometimes are not observed at all, \eg, GRB061007, \citep{Panaitescu2007,Racusin09}.\\

  {\bf Correlated X-ray-GeV signal}
Detection of GeV photons \citep[\eg\ in GRB080916C,][]{FermiGRB080916C} requires large emission radii, $r_{GeV} \sim 10^{16}$ cm. The requirement of low optical depth and conventional scaling of variability, $ r \sim \delta t_v \Gamma^2$,
 then imply immense bulk Lorentz factors  $\Gamma \sim 1000-2000$. Note, that in a standard  fireball  model, there is an upper limit on bulk Lorentz factor, 
 $\Gamma \leq 1000 \, L_{51}^{1/4}\, r_{0,6} ^{-1/4}$ \citep{2005ApJ...635..516N}, where $r_0$ is the initial radius of the outflow.  In case of larger $\Gamma$, corresponding to  less  baryon-contaminated flows, most of the flow energy remains in the form of thermal photons, and not in the bulk motion of baryons.  The Lorentz factors inferred for {\it Fermi} GRBs come close to violating this limit. The electromagnetic model \citep{LyutikovJPh}, which typically has larger Lorentz factors than the fireball model,  does not suffer from the upper limit on Lorentz factor, see also Fig. \ref{xiless1}.

Whether the prompt and the GeV emission come from the same region is an open question. On the one hand, correlated variability and a  single spectrum in case of GRB080916C argues for co-spacial origin. On the other hand, GeV emission continues for hundreds of  times longer than the prompt, and separate spectral component is clearly seen in GRB090510 and  GRB090902B, see \S \ref{Fermisynchrotron}.

{\bf  Similarity of Short's and Long's very high energy emission}.
The latest Fermi results \citep[\eg\ on GRB080916C][]{FermiGRB080916C} make the similarity between the Shorts' and Longs' long lasting emission  even more surprising. GeV signals from both types of bursts start with a short delay, of fewer than several seconds,  with respect to prompt emission and continues for hundreds to thousands seconds after the prompt emission.  Such continuity, from nearly simultaneous with prompt to long lasting emission in both cases is quite  surprising, especially  if photons are produced in the forward shock with drastically different properties for Longs and Shorts. 

 {\bf Missing reverse shocks}.
The standard model had a clear prediction, of a bright optical flare with a definite decay properties \citep{1996ApJ...473..204S,1997ApJ...476..232M}. Though a flare closely resembling the predictions was indeed observed \citep[GRB990123,][]{1999MNRAS.306L..39M}, this was an exception.  In the {\it Swift} era,  optical flashes are rare \citep{Gomboc};   even when they are seen, they rarely correspond to RS predictions (often optical emission correlates with X-rays,  too variable, wrong decay laws).

\begin{figure}[h!]
 \begin{center}
\includegraphics[width=\linewidth]{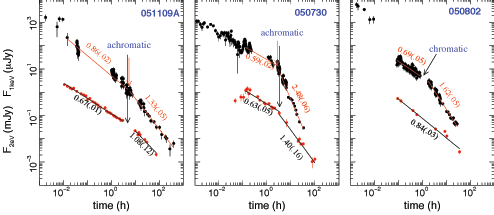}
\end{center}
\caption{ Examples of light curves showing achromatic breaks, breaks  with different jumps in optical and X-rays, chromatic breaks \citep{Panaitescu2007} }
\label{JetBreaks}
\end{figure}

{\bf Missing jet breaks}.
The FS model of afterglows predicts, that when the bulk Lorentz factor becomes  $\Gamma \sim 1/\theta_j$ ($\theta_j$ is jet opening angle), an achromatic break should be observed in the light curve \citep{Rhoads}.
There are examples of GRBs with missing break \citep[\eg\ in  GRB061007 the B band optical light curve is fitted with a  well sampled power law between $80$ and $10^6$ seconds,][] {Racusin09}. In different  cases the  breaks can be  achromatic, with different jumps in optical and X-rays, and  chromatic, Fig. \ref{JetBreaks}  \citep{Panaitescu2007}.

The problem of missing jet breaks puts a question mark on the use of GRBs as cosmological probes. A number of relations involving jet break time were proposed
 \citep[\eg][]{Frail01,2004ApJ...616..331G},
 which lead to identification of a ''GRB standard candle", and possible routes to probe geometry of the Universe.  
These correlations are now in doubt. 
  Note, that $E_p - L_{iso}$ - correlation discussed above in application to the synchrotron model, involves the  correlations between local quantities, which can more  naturally be expected, while the 
 $ E_p$ - jet angle correlations (e.g. Ghirlanda correlation) involve  local and  global quantities.


\section{ Fermi   problem with synchrotron}
\label{Fermisynchrotron}
Previously, 
a number of authors discussed the detection of high energy emission from GRBs by EGRET \citep{Gonzalez03,2006NJPh....8..122D,2005ApJ...618L..13B}. By far the most common explanation was (is) the IC emission by electrons accelerated either  internally or at the FS. Detection by {\it Fermi} of 13 GeV photon (nearly 50 GeV in rest frame) from GRB080916C seemed at first to confirm these expectation. Yet, surprisingly, the whole spectrum, from 8 keV to tens of GeV is fit with a single power law. In case of  SSC-type model (synchrotron for X-rays and  IC for  100 MeV - GeV photons) this requires fine-tuning, so that the IC spectrum would match the lower energy synchrotron component \citep[see, though][]{2009arXiv0905.2417K}.
Note also, that the two other well sampled bursts  detected by {\it Fermi} in the GeV range, GRB090902B and  GRB090510, do show a different component in the GeV range. 

In fact GeV emission, internal or external,  almost surely does not come from synchrotron emission by particles accelerated via stochastic Fermi process. Radiation-limited acceleration cannot produce synchrotron photons with energy higher than about $30 $ MeV in the rest frame. This limit comes from balancing the most efficient acceleration, by electric field equal to \Bf\ \citep[in case of stochastic acceleration models, this is also equivalent to scattering on relativistic cyclotron time scale of a particle][]{1996ApJ...457..253D}:
\ba &&
e E c \sim e B c  \approx {e^2 \over c } \om_B^2 \gamma^2
\nn &&
\om \sim \om_B \gamma^2,  \, \rightarrow \, \lambda = {e^2 \over m_e c^2}, \epsilon \approx 30 {\rm MeV}
\label{lambda}
\ea
Since transparency of the outflow demands that bulk Lorentz factor be $\Gamma\sim 1000$, this limit is barely reached for GRB080916C. But we stress that this is an absolute upper  limit that assumes non-stochastic acceleration. Modern simulations \citep{SironiSpitkovsky09} indicate that the limit of non-stochastic  acceleration can, in principle, be reached in drift acceleration, but this requires a very particular shock geometry (nearly parallel shocks in case of relativistic shocks).  A possible caveat  is inhomogeneous \Bf: particles are accelerated in low field regime, and emit synchrotron radiation in regions of high \Bf. 
Thus, GeV photons cannot be generated by synchrotron mechanism, regardless of whether they are coming from internal dissipation or external shock \citep[\eg, as suggested by ][]{2009arXiv0905.2417K}. 

A possibility that GeV photons come from baryonic processes must be mentioned. At $\Gamma \sim 1000$, protons and neutrons have TeV energies, so that inelastic collisions or just neutron decay will generally produce $\sim 100 $ GeV photons. The problem with this interpretation is that $p-n$ collisions are frequent enough only until 
$r \leq 10^{13}$ cm, while transparency requires $r \approx 10^{16}$ cm. In addition, this is a relatively inefficient way to generate high energy photons. 
On the other hand, if the constraints of the hadronic models for GeV emission can be overcome, this will be an argument in favor of acceleration of UHECRs by GRBs. 
A complication, though, is that  in stochastic acceleration models it would be   required that efficient, nearly Bohm-type,   scattering by turbulence extends from  scales corresponding to  $\sim    100 $ GeV electrons to $ 10^{20}  $eV ion gyroradii, since for less efficient acceleration the adiabatic losses of CRs would dominate.

\section{Possible resolutions: X-ray afterglows from internal dissipation}

\subsection{FS is there, but is it seen at early times?}

The paradigm of external-reverse shocks in astrophysical outflows is well established, primarily by observation of SNe \citep[\eg][]{Reynolds08}.  (Recall, though, that the FS is {\it not} observed from, perhaps, the most studied object in high energy astrophysics, the Crab nebular.)
In case of GRBs, two issues may complicate the picture. First, the FS is strongly relativistic, at least at early times. It is not clear, at least observationally,  if relativistic shocks can accelerate particles as efficiently as non-relativistic one \citep{SironiSpitkovsky09}. The FS must be there, but it might be unobservable, at least at the relativistic stage. Secondly, GRB outflows can be magnetically dominated \citep{LyutikovJPh}; this can suppress reverse shock emission.

In the pre-{\it Swift} era, late afterglows, at times   $\geq 10^5$ seconds, preferentially showed smooth light curves, with power-law decays, generally consistent with the  emission of a FS. Note, that these observations were mostly done in the optical and  radio waveband.
 Now, in the {\it Swift}
era,
 early, $\leq 10^5$ seconds, broad-band observations of GRBs, most importantly, done in X-rays,  often show numerous features in a form of flares, plateaus, various light curve breaks. This variability runs contrary to the predictions of the FS model, which predicts a smooth light curve, with one possible achromatic jet break. 
 Most importantly, this variability often is uncorrelated  between the  X-ray and the optical wave wave bands. 
Below we discuss a possibility that  {\it conventional X-ray and partly optical GRB afterglows, especially at early  times $\leq 10^5$ sec, might be coming from internal dissipation, and not from the forward shock} \citep[this type of models were recently advocated by][]{Uhm,Genet,2008MNRAS.388.1729K,CannizzoGehrels}.

 Our knowledge of the  non-thermal signal from  SNe associated with GRBs is dominated by one particular event, GRB030329, associated with 
  SN 2003dh. 
  Other famous SN-GRB connection, GRB980425-SN1998bw   \citep{Galama98} and GRB031203-SN2003lw \citep{2004Natur.430..648S} 
   involved  sub-energetic GRBs. 
   In  {\it a few} cases where SN signature in clearly seen in GRBs, there is a clear {\it separate } radio component appearing after approximately $1$ day in radio and sometimes in optical, which then peaks at $\sim 40$ days (in case of GRB030329) and  continues for months and years \citep{Berger03}. 
 This is clearly the FS:  the expansion of the radio afterglow from few days onward matched the prediction of a blast wave model \citep{Taylor04}; after almost ten years, the evolution of the radio afterglow is still well described by the expansion \cite{2005ApJ...619..994F}. 
Overall, this is what is expected in radio band:  radio  emission is
self-absorbed at early times the peak synchrotron frequency has not yet cascaded
to the radio band. 

On the other hand, in X-rays, there is also a new component appearing at $\sim 40$ days after the burst \citep{Tiengo}, approximately around time when the FS becomes non-relativistic. 
 Thus, the fact that FS shock is not seen in X-rays at earlier times, $t \leq 10^5 $ sec. does not contradict observations. 
 Forward shock eventually must become visible, at least in a non-relativistic phase: we know from observations of SNRs, that in the non-relativistic regime the FS produces a clear  non-thermal signatures, as indicated by radio emission and   non-thermal X-ray filaments observed in young SNRs \citep{Reynolds08}. Note, though, that in case of SNRs, 
 the amount of energy emitted by the FS through  non-thermal emission is typically small compared with the energy budget and with the thermal emission 
 (most thermal emission is coming from the reverse shock, though).

\subsection{Extended activity in Longs: failed SN}

Modern models  of neutrino-driven SN explosion are not 
ÒstableÓ, in a sense that  different groups do not 
agree with each other and the role of different ingredient is not settled. 
In addition, the role of \Bfs\ is not clear  \citep[the idea of magnetically-driven explosions dates back to][]{LeBlancWilson,Bisnovatyi71}: in {\it all}  present-day  simulations of a  successful magnetic 
explosions, a  very high 
initial \Bf\  is needed \citep{2007ApJ...664..416B,2009arXiv0907.0561N}. It is feasible that such field are indeed achieved in nature, but testing this paradigm requires full 3-D relativistic MHD simulations of magnetic dynamo; something  which has not been done yet for this problem.  Thus, the
initial conditions for 
the GRB engine are not clear. 

It is possible that depending on the detailed properties  of the pre-collapse core (like angular momentum, initial \Bf, small differences in composition etc), the
two energy sources  that may potentially lead to the explosion, neutrino-driven convection and Faraday-wheel type dynamics of \Bfs, may contribute different amount of energy, resulting in different observed types of SNe and/or GRBs. Neutrino-driven explosions generates  quasi-spherical sub-relativistic outflows, while magnetic-driven explosion results in a  jetted relativistic component. For example,  classical SNe are all neutrino driven, while  magnetic engine is  negligible. 
As the  relative strength of the magnetic engine increases, this leads to phenomena of sub-energetic and regular GRBs. The relative contribution of the 
neutrino-driven and rotational energies extracted by \Bf\ are definitely not independent of each other: one expects that in case of a  successful neutrino-driven SN, the amount of material accreted on the central source is small, resulting in weak or no relativistic component and a weak GRB:  recall that all SN-associated GRBs are subluminous.



\subsection{Late accretion from a failed SN}
\label{failedSN}
Production of  X-ray afterglows via internal dissipation (as opposed to the external forwards shock), offers a number of qualitative advantages, especially in explaining 
 plateaus and early  flares.  In case of internal dissipation, the emitted power is related to the instantaneous power of the central source.
In contrast, in case of   FSs, the  emitted power is related to the {\it total} energy deposited into ICM. Thus, in order to modify emission properties of a FS, one needs to inject energy comparable to the total energy injected previously. 
Though both FS and internal dissipation model do require long lasting engine ({\it at least $\sim 10^4 $ sec} and longer,  the energy requirements  are drastically different, by at least an order of magnitude: FS requires that the late activity contains ten to hundred time more energy than the prompt emission, while the internal dissipation requires comparable contribution. Note also, that 
plateaus  are seen only in Longs, consistent with the accretion of an extended
 envelope in a failed SN.  Envelope  accretion  time scale, $t_{acc} \sim 1/\sqrt{G\rho} \sim 10^3 -10^4$ sec, 
 also agrees with the expectations from the structure of a pre-SNib/c stars.
 
 There are two possibilities for the internal production of early X-ray afterglow.  \cite{Genet,Uhm}  suggest that emission is coming from the reverse shock. By tuning the 
 luminosity of the central engine as a function of time, one can reproduce a  variety of behaviors, including plateaus, flares and late jet-breaks. 
  A drastic difference with the standard picture is that the RS particles are in the fast cooling regime
     
   Alternatively, 
\cite{2008MNRAS.388.1729K,CannizzoGehrels} proposed that early afterglows may come from the accretion disk (or the accretion disk corona) formed around the central source. 
 The advantage of these models is that
 the variability time scale is different for optical and X-rays, since they are coming from different parts of the disk (outer and inner, correspondingly). This naturally explains the achromatic breaks. Disk instabilities, well monitored in Galactic X-ray binaries, 
 induce variation of disk properties on accretion time scales. 
 In case of internal dissipation, a flare corresponds to a change of {\it rate of energy dissipation} (again, in case of a  FS,  a flare corresponds to a change of {\it total  energy } of outflow). Each  flare does not contribute much to a total energy budget of the outflow. 

 The proposal that X-ray afterglows are coming from late envelope accretion bears a clear prediction: there should be no plateaus and/or powerful early time flares in GRBs with a clear SN signature. The GRB030329 complies with this prediction: it dod not show flaring.  Also, plateau phases are not seen in Shorts, again consistent with the envelope accretion in failed SNe. 
 
 Both types of proposals, that the  early X-ray afterglows coming from internal dissipation in the reverse shock or the accretion disk, suffers from difficulties, though. If emission is coming from the RS as suggested by   \cite{Genet,Uhm}, then it is hard to reproduce sudden drops after the plateaus \citep[][Fig. \ref{Troja}]{Troja07}; in addition, since RS is indicative of interaction with external medium, the proposal suffers from the same difficulty in explaining why afterglow properties are strongly correlated with the prompt \citep[][\S \ref{Problems}]{Nysewander}. In addition, at the end of the plateau, the spectrum remains continuous over the break \cite{Nousek06}.  Also,  even though sometimes plateau phases end  with an abrupt break, most of the time  there is a smooth, continuous transition. 
Both these facts  indicate that the plateaus and post-plateau emission are coming from the same component, whose properties are changing, and not from two different overlaid components.  This implies that {\it all  X-ray afterglows, up to $10^5-10^6$ sec  also come from internal dissipation. } 
 
  If early X-ray afterglows are coming from the disk emission \citep{2008MNRAS.388.1729K,CannizzoGehrels}, it is easy to reproduce sudden drop-offs of emission at the end of the plateau phase.  In case of disk accretion  model,  no relation between spectral and temporal indexes is expected (one is determined by the temporal evolution of the power of the central source, another by the local particle accelerated spectrum). In addition, since behavior of inner and outer parts of the disk occurs on different times scales, one naturally expects achromatic behavior of light curves.   The main problem of the disk accretion models is the similarities between afterglows of Longs and Shorts: virtually in any model of Shorts, a long lasting $t \geq 100 $ sec, accretion is not expected. 
 

Thus,  the early X-ray afterglow emission can be  attributed to internal dissipation of the flow: in the RS \citep{Genet,Uhm} or in the disk Corona \citep{2008MNRAS.388.1729K,CannizzoGehrels}. But where is the FS shock emission then? FS can still appear in X-rays, but at a much later times (\eg\ at 40 days in case of GRB030329).

\subsection{Magnetar-type central engine} 
\label{23}

An alternative possibility for the internal production of X-ray afterglows, is a magnetar-type central engine. This has a potential advantage to explain similarities between Longs and Shorts, as being due to the same central engine. In case of Longs,  magnetar-type central engine is a long-standing alternative to the collapsar model. 
In case of Shorts, magnetar-type object may form in a hypothetical accretion induced collapse  (AIC) of a white dwarf. Note, that in this case the collapse itself takes 
$t_c \sim 1/\sqrt{ G \rho_{WD}} \sim $ tens of seconds, yet the dissipative stage is much shorter, corresponding to the density at the time of the bounce, at  nearly \NS\ radius; this is of the order of a fraction of a second  \citep{Fryer99,2009arXiv0910.2703A}. The main problem in associating the Shorts with AIC is that at the time of the bounce, the flow is expected to be very hot and as a result too  strongly baryon loaded \citep{2007ApJ...669..585D}.

In case of a magnetar-type central engine, one expects a long-lasting activity, decaying according to a prescribed energy loss mechanism (\eg\ dipolar spindown).  
One expects a break at time 
\be
t_b \sim { I c^3 \over B^2 R_{NS}^{6} \Omega^2} \sim 10^4\, {\rm sec} B_{15}^{-2} P_{-2} ^2
\label{tb}
\ee 
($I$ is moment of inertia of a \NS\ $\Omega $ is rotation frequency, $P$ is the spin period), 
corresponding to a spin-down time of a \NS\ with surface \Bf\ $B$ and initial spin $\Omega$. 

The spin down time (\ref{tb})  has a strong dependence on the surface \Bf\ and spin period, and can last from $\sim $ hundred  seconds to much longer. The time $t_{b} $ can be related to the plateau phase in Longs and extended tails in Shorts.   In  the extreme case of millisecond period and super-critical \Bfs, this will give a short-lasting, powerful signal, which energetically may even be dominant over the prompt, like in  GRB080503 \citep{Perley}.
For smaller \Bfs\ and/or longer periods, spin-down luminosity is small, and would fall below the detection level. Thus, this model predicts that all SGRBs have an extended emission period.

Since the magnetic-driven outflows are expected to be strongly anisotropic, one expects that in case of barely-failed SN, the collimated jet breaks out along the polar directions, while in the equatorial plane material accretes on the central engine. The estimates for the envelope accretion in failed SNe, \S 
\ref{failedSN}, are applicable to the magnetar central engine as well:
the time scale of such accretion will be 
$t_{acc} \sim 1/\sqrt{G\rho} \sim 10^3 -10^4$ sec, where $\rho \sim 10-10^3 $ g cm$^{-3}$ is a typical density of the outer shell in the SN Ib/c progenitors.
This time scale may be identified with plateaus in Longs (NB: there are no plateaus in Shorts). In addition, powerful frequent flares {\it at the  plateau phase}
may be due to non-stationary accretion (\eg, in X-ray binaries we observe disk instabilities that lead to a variability on a broad range of time scales). 
Naked  GRBs (GRBs without an appreciable afterglow emission) correspond to weak extended outflow (low {\Bf}s and/or long initial periods).

We propose that  there are two flare production mechanisms. First operates only in Longs, producing powerful frequent flares {\it at the  plateau phase}, and is due to non-stationary accretion of the outer layers of  a failed SN. The second types of flares, much less frequent and less energetic, are  produced in  Shorts and   in Longs at later times, after the end of the  plateau phase.  They can be identified with magnetar flares  driven by rearrangement of super-critical \Bf\ in the newly born \NS \citep{palmer,tlk}.

 \begin{wrapfigure}{l}{0.49\textwidth}
 \vskip -.3 truein
 \begin{center}
\includegraphics[width=.99\linewidth]{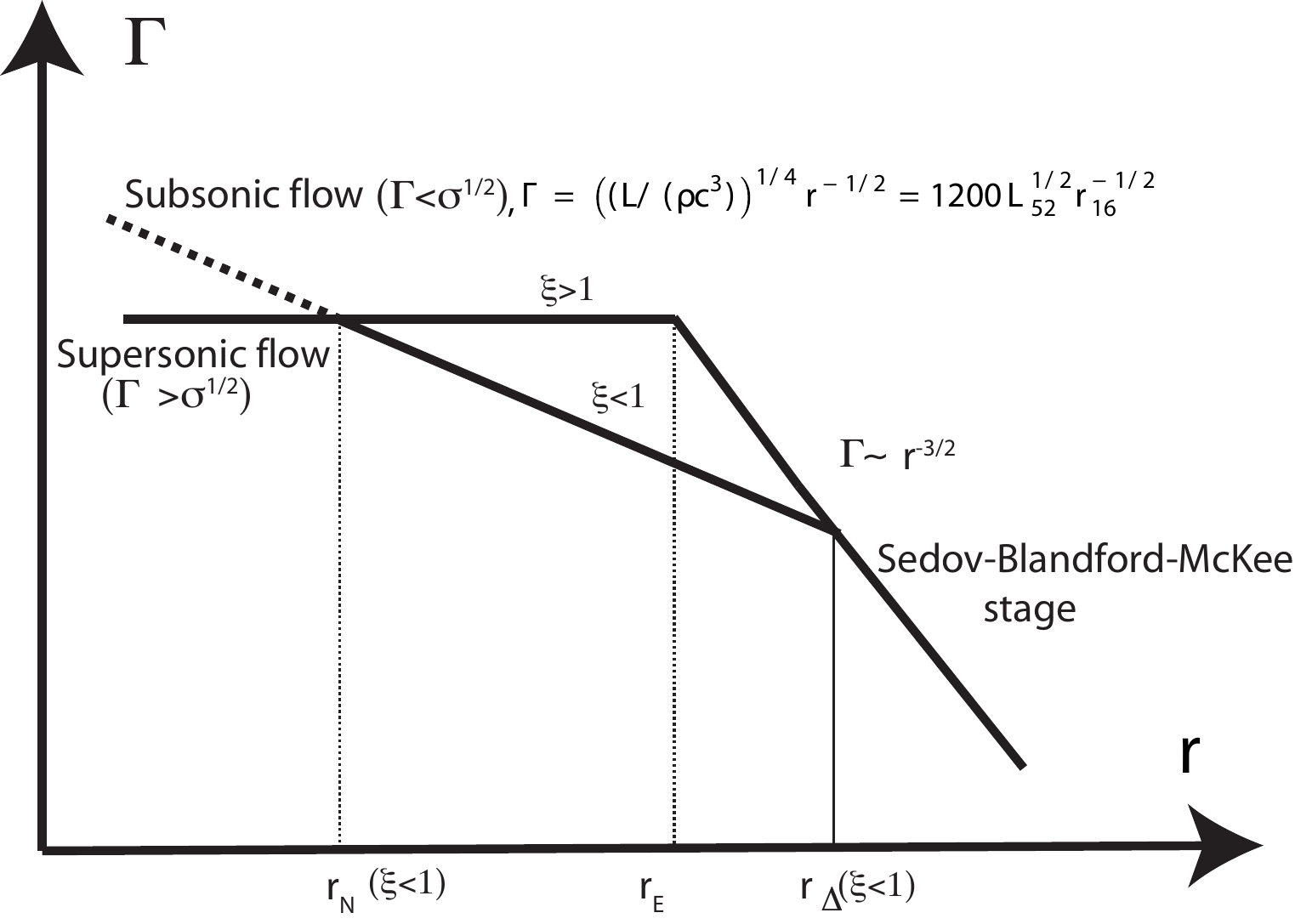}
\end{center}
 \vskip -.2 truein
\caption{Evolution of the Lorentz factor of the outflow for the  fireball (solid line) and electromagnetic (dashed line) models. For subsonic ejecta, $\Gamma < \sqrt{\sigma}$, the  Lorentz factor initially decreases $\Gamma_{ISM}\propto r^{-1/2}$, changing to Sedov-Blandford-McKee regime $\Gamma_{ISM}\propto r^{-3/2}$ at $r_\Delta$. For supersonic ejecta with initial Lorentz factor $\Gamma_w$, the flow either coasts with $\Gamma= \Gamma_w$ until $r_E$ (for $\xi>1$, so called thin shell, \cite{Sari97}), or has an intermediate regime $r_N < R< r_\Delta$ where $\Gamma\propto r^{-1/2}$ (for $\xi<1$, thick shell). Parameters are
$\xi=   \sqrt{ l_S \over  \Delta} \Gamma_w^{-4/3}, \, r_\Delta  = l_S ^{3/4} \Delta ^{1/4}, \,  r_N = {l_s ^{3/2} \over  \gamma_4^2 \Delta ^{1/2}} ,\, r_{E}  = {l_S \over \Gamma_w^{2/3}}, \, l_S = \left( { E_{\rm iso} \over  \rho c^2  } \right)^{1/3}$. }
 \vskip -.2 truein
\label{xiless1}
 \vskip -.25 truein
\end{wrapfigure}    
The main advantage of the magnetar-type central engine is that it can provide a long lasting central engine, behaving similarly in both Longs and Shorts. 
The main problem with early afterglows coming from internal dissipation in a magnetar-drive outflow is  the emission mechanism. If it is similar to the prompt emission, then why prompt is so variable, while afterglow are much smoother (this can be expected if the  outflow becomes much less relativistic later in time). If the afterglow emission is from RS, then, again,  why afterglows from Longs and Shorts look similar?

\section{The role of \Bfs}

The role that the  \Bfs\ play in  acceleration, collimations and 
  emission production in GRB outflows remains one of the central issues.  The fireball model advocates that 
  magnetic fields do not play any major dynamical role (except, perhaps, at a very early stage,
after which  fields are quickly dissipated).
In the emission region magnetic field are
 re-created locally 
   with energy density typically much smaller than plasma energy density. Fields are
   small scale, with correlation length $l_c$ much smaller
   than the "horizon" 
    length, $l_c \ll R/\Gamma$.
 Alternative approach, advocated by MHD and electromagnetic models \cite[\eg][]{Usov92,lb03,Vlahakis03,LyutikovJPh}
  is that there are dynamically important
   large scale fields with "super-horizon" correlation length $l_c \geq R/\Gamma$, which are 
    created at the source and which may play a major role in driving the
    whole outflow in the first place.

 \begin{wrapfigure}{l}{0.5\textwidth}
 \vskip -.2 truein
\includegraphics[width=0.95\linewidth]{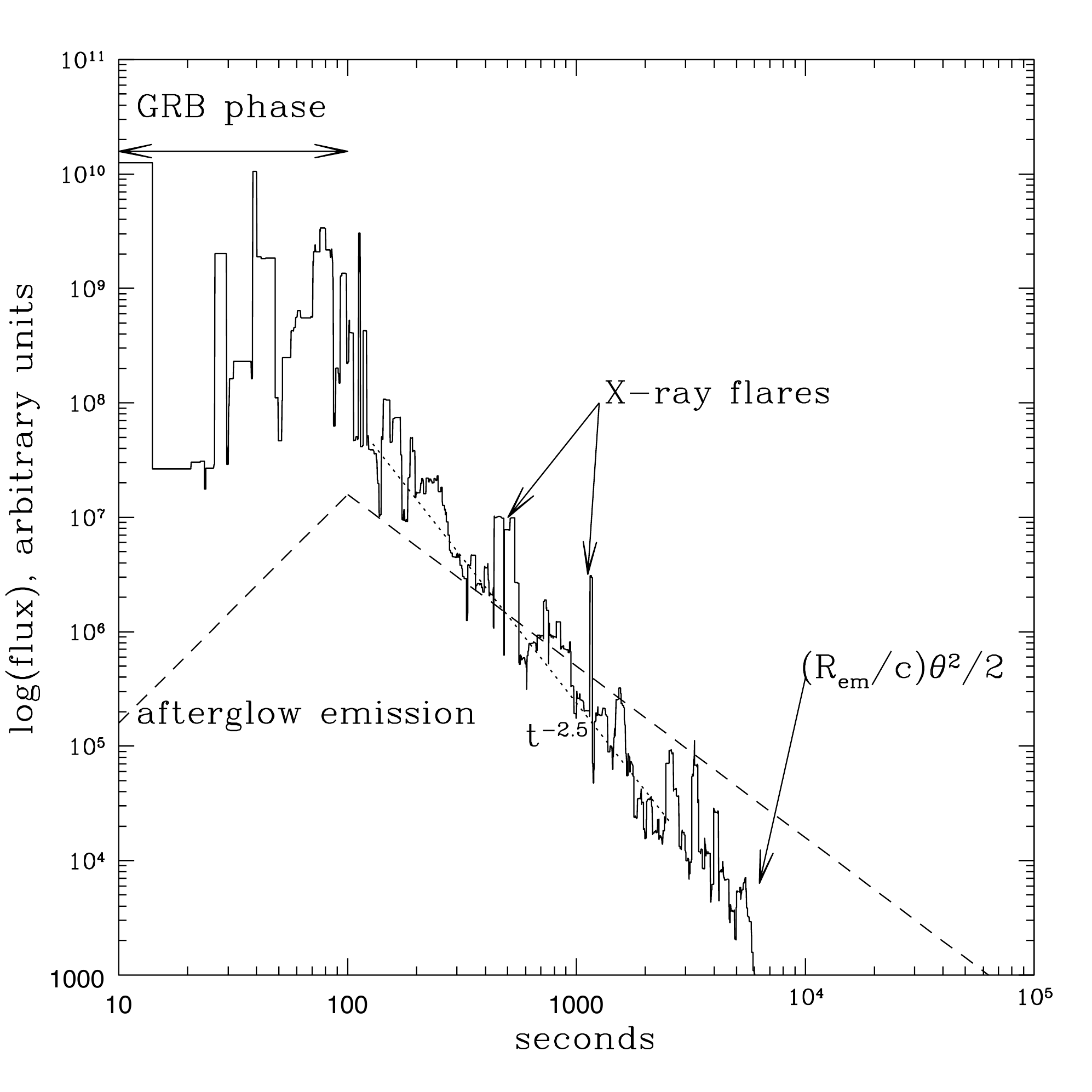}
\caption{Prompt and early afterglow emission produced by emitters  moving  randomly  in the bulk frame \cite{LyutikovRem}.}
 \vskip -.2 truein
\label{GRBafter}
\end{wrapfigure}       
         Within the framework of  electromagnetic models, $\gamma$-ray-emitting electrons are accelerated at distances  $r \sim 10^{15}- 10^{16} $ cm 
 due to development of electromagnetic current-driven  instabilities and ensuing formation of reconnection layers.   No shocks are needed.  The conversion of magnetic energy
    into particles may be very efficient.
    For example, {\it RHESSI } observations of the Sun indicate that, in reconnection
    regions, most magnetic energy goes into non-thermal electrons with power-law distribution in energies
    \citep[][]{benz03}. Though the details of how 
   magnetic dissipation and particle acceleration
    proceeds in Solar flare and  GRB outflows are bound to be different, the underlying principles may be
    similar.  In particular, in the electromagnetic model the variability of the prompt emission reflects 
the statistical properties of dissipation and not the source activity as in the fireball model. 
Magnetic fields are non-linear dissipative  dynamical system which often show 
bursty behavior with 
power law  PDF.
For example,
solar flares  show variability on a wide
range of temporal scales, down to minutes, 
which are  unrelated to the time scale of 22 years of magnetic field generation in the tachocline.

    Are there evidence for  presence of large-scale \Bfs\ in GRB outflows? After the initial claims of high prompt polarization \citep{CoburnBoggs} and ensuing rebuttal
     \citep{RutledgeGRB}, the evidence are inconclusive. A recent detection of high optical polarization from what appears to be a RS (Steele \etal), is intriguing, but requires an independent confirmation. 
    Even if large scale \Bfs\ are present and produce high polarization, this still does not imply that they are dynamically dominant.

     An argument in favor of energetically dominant \Bfs\ come from the interpretation of early plateaus and flares as the products of late time central engine activity. 
    The energy deposition rate  via the neutrino annihilation process
drops precipitously below the accretion rate of $\sim 0.02 M_\odot sec^{-1}$
    \citep{1999ApJ...518..356P} (black hole  spin  increases somewhat  the efficiency of neutrino annihilation \cite{2008AIPC.1054...51B}). Thus, long engine activity requires magnetic driving \citep{Barkov08}, \eg\ through Blandfordf-Znajek process \citep{BlandfordZnajek}.
     
    The {\it Fermi} observations of prompt GeV photons and corresponding requirement of high Lorentz factors, $\Gamma \sim 1000$ may also be interpreted as an indication of dominant  
        \Bfs. For highly magnetized flows, the Lorentz factor of the outflow is similarly high: $\Gamma \sim \left( (L/( \rho c^3) \right)^{1/4} r^{-1/2} = 1200 L_{52}^{1/2} r_{16} ^{-1/2}$
        \citep[Fig. \ref{xiless1}, see also][]{LyutikovJPh}. 
 
 \subsection{Fast variability from large radii: mini-jets}

If prompt emission is produced at distances $\sim 10^{15}-10^{16}$ cm,
how can fast variability,
on times scales as short as milliseconds, be 
achieved? One possibility, is that emission is beamed in the outflow frame, for example 
due to 
 relativistic motion of ''fundamental emitters'' \citep{LyutikovRem}.
As was shown by \cite{LyutikovRem} (see also \cite{2009MNRAS.395L..29G,2008MNRAS.386L..28G,2009MNRAS.395..472K},  Fig. \ref{GRBafter}),  a highly variable, efficient prompt emission can be produced in the framework of the model. 

The key point is that in case of relativistic random motion with random  Lorentz factors $\gamma_{rand} $, the effective Lorentz factor becomes $\Gamma_{eff} \approx 2 \Gamma \gamma_{rand} $. Thus, even if $\gamma_{rand} \approx 2$, the temporal variability becomes shorter by a factor $4  \gamma_{rand}^2 =16$, while the observed flux during a time that the random emitter is ''looking'' at the observer increases by an even higher factor. As a result, emission from large radii can be highly variable, and yet sufficiently effective, tapping into a large proper volume \citep{LyutikovRem}. In relation to GeV photons, if the seed photons are limited to the moving blob, while the inter-blob medium is devoid of photons, this also alleviates the problem of large required  bulk Lorentz factors. 

Possible origin of relativistic motion of ''fundamental emitters'' may be the fact that in case
of relativistic reconnection occurring in plasma  with $\sigma \gg 1$,
where
$\sigma$  is a plasma magnetization parameter \citep{kc84},  the out-flowing matter reaches
relativistic speeds with $\gamma_{out} \sim \sigma$ \cite[]{LyutikovUzdensky}.
Internal synchrotron emission by such jets,
or  Compton scattering of ambient photons, will  be strongly beamed in the frame of the outflow \cite{LyutikovRem,2009MNRAS.395..472K}. 

 \begin{wrapfigure}{l}{0.5\textwidth}
\includegraphics[width=0.95\linewidth]{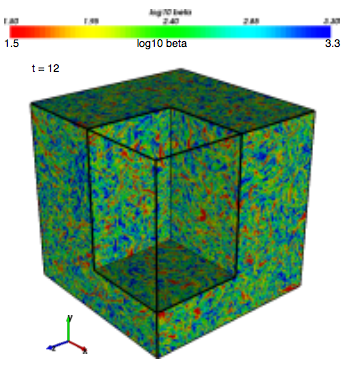}
\caption{Relativistic MHD  turbulence \citep{ZhangMacFadyen}. The picture shows plasma $\beta$, the ratio of gas pressure to the magnetic pressure. In a developed turbulent state plasma beta is 
highly variable, showing formation of current layers. Simulations are  sufficiently resistive, though, so  the  current sheets do not form. Is it expected that  for smaller resistivity,  current sheets and reconnection layers will indeed form. }
\label{MacFadyen-Turbulence}
\end{wrapfigure}
Reconnection layers can form as  a result of relativistic magnetized turbulence   \citep[][see  Fig. \ref{MacFadyen-Turbulence}]{ZhangMacFadyen}. \footnote{In passing we note, that modifications of this model based on an idea of fluid relativistic turbulence \citep{PiranTurb}  are not physically justified. A description of turbulence as a collection of fluid vortices is a very specific model of  incompressible fluid turbulence and is not applicable to relativistic GRB outflows. On basic grounds, one expects that random motions will have speeds smaller than the sound speed, giving random Lorentz factors  not much different from unity, $\gamma_{rand} \leq \sqrt{ 9/8} =1.06$. }
What triggers the reconnection? Qualitatively, the lower the plasma density, the more resistive it becomes. 
One possibility is that plasma density becomes  so small,  that plasma cannot support internal currents \citep{Coroniti90,LyutikovBlackman}. Another possibility is that the rate of reconnection changes from slow  Sweet-Parker rate to fast  Petschek rate, when the plasma becomes  sufficiently tenuous and collisionless \citep{Uzdensky07}.

The mini-jet model has a number of clear predictions: spectrum should be harder during the flare; this seem to agree with observations (Burrows et al 2005); flares  should become   longer and softer as function of flare time. 
Can some  Shorts be Òone spike LongÓ, of a failed SN type? Also this model may, perhaps, explain the correlated  optical-$\gamma$-ray variability in GRB080319B,  \eg\ if  emitting ÒblobsÓ expand, killing both synchrotron and IC luminosities. 

\section{Instead of conclusion}

Our lack of understanding of how GRBs work is impeding the progress and applications of GRBs, \eg, to cosmology.  One can only hope that GRB theory will not
suffer the same destiny as theories of pulsar radio emission: though we do not understand it and basically gave up hopes, pulsars are still very useful probes of interstellar medium and General Relativity. In case of pulsar radio emission, only about one millionth part of energy is emitted as radio waves; this make the problem especially hard. 
In contrast, in case of GRBs we are dealing with the dominant energy release, of the order of Solar rest mass energy in a matter of seconds. It is frustrating that we have a hard time to understand even the basic elements in such energetic phenomena. 

I am greatly indebted to many colleagues for stimulating discussions and the organizers of the meeting ''The Shocking Univers" for hospitality. In particular, 
I would like to thank Andrei Beloborodov,  Guido Chincarini, Dale Frail, Jonathan Granot,  Pawan Kumar, Ehud Nakar,  Alicia Soderberg and  Bing Zhang for comments on the manuscript.

\bibliographystyle{apj}
\bibliography{/Users/maximlyutikov/Home/Research/BibTex}

\end{document}